# Detection of Colluded Black-hole and Grey-hole attacks in Cloud Computing

Divyasree I R., Selvamani K., Riasudheen H

*Abstract* -The availability of the high-capacity network, massive storage, hardware virtualization, utility computing, service-oriented architecture leads to high accessibility of cloud computing. The extensive usage of cloud resources causes oodles of security controversies. Black-hole & Gray-hole attacks are the notable cloud network defenseless attacks while they launched easily but difficult to detect. This research work focuses on proposing an efficient integrated detection method for individual and collusion attacks in cloud computing. In the individual attack detection phase, the forwarding ratio metric is used for differentiating the malicious node and normal nodes. In the collusion attack detection phase, the malicious nodes are manipulated the encounter records for escaping the detection process. To overcome this user, fake encounters are examined along with appearance frequency, and the number of messages exploits abnormal patterns. The simulation results shown in this proposed system detect with better accuracy.

## I. INTRODUCTION

Cloud Computing is a robust- connected predominant and on-demand technology beneficial for business services. It provides better result such as scalability, flexibility, capacity utilization, and mobility, cost-effective and higher efficiencies. The major constituent of cloud computing is software architecture, hardware resource that enables virtualization and scalable infrastructure. The cloud architecture provides measured cloud services tendered by the cloud service provider to cloud consumers over a networked infrastructure. Even though, the cloud computing solves difficult problems in IT industries but also faces a new threat to the data in terms of security. The significant security risk in the cloud environment is data security, logical access, physical security, Network security, virtualization issues and compliance [5].

The most common and simple way of protecting a network resource is by assigning it to a unique name and its corresponding password. Cloud Computing is threatened by various attacks, including Denial of service (blackhole & greyhole attacks), Distributed Denial of Service Attack (DDoS) [2], Cross-site scripting, Network Sniffing etc. Black-Hole (False Report) attack & Gray-hole (Packet Drop) attacks are dangerous to cloud network attacks. A malicious node with enough buffer storage misleadingly drops or routes the message packet on the way which consumes more energy is referred to as Black-Hole attack. The Black-Hole Node (BHN) acknowledged without any verification whether it has the correct path or not [3, 6]. This forwarded packets which are dropped is shown in figure 1. In a selected portion of message packets, a packet drop for every 't' seconds or every *'n'* packets for a particular specified network destination is referred to as Gray-Hole attack.

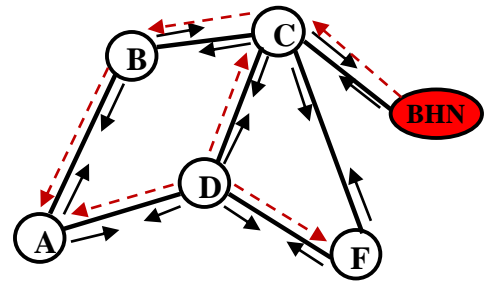

Figure 1: Black-Hole Attack.

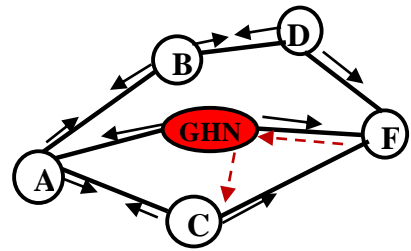

Figure 2: Gray-Hole Attack.

The figure 2 depicts that source node A sends a message to destination node F, where Gray-Hole Node(GHN) drop the route reply and packets are forwarded by node D to A[9,10]. The dropping misbehavior will decrease the overall message delivery and waste the resources of intermediate nodes that have carried and forwarded the dropped messages. Dynamic trust management protocol can deal with selfish behavior and is resilient against trust related attacks.

The rest of the paper is structured as follows: The existing detection methods against Black-Hole & Gray-Hole are summarized in section 2. Section 3 describes the proposed system for individual and collusion attack in cloud computing. Section 4 explains the evaluation parameters and simulation result. In section 5, concluded the proposed work and suggested some possible future works.

## II. LITERATURE SURVEY

The characteristics of Black-Hole & Gray-Hole attacks that decreases the message distribution by dropping the messages which are taken by the intermediate node with excess resources.

*Thi Ngo et al.* [12] designed an integrated detection system for both individual and collection attack misbehaviors. The detection algorithm is referred to as Statistical based Detection of Blackhole, and Greyhole attackers (SDBG) and forwarding ratio metrics categorize the attacker's misbehaviors from normal nodes. Ferry-Based Intrusion Detection and Mitigation (FBIDM) [1] techniques were proposed for network attack detection. The trusted ferries used to check whether the nodes increase their delivery probabilities to absorb more data. This scheme is auspicious in dropping the impact of malicious attacks.

*Ren et al.* [8] discussed a Mutual Correlation (MUTON) detection scheme for insider attacks in the network. When the calculation of packet delivery probability of node MUTON considers the transitive properties and collected information are correlated with other nodes. The compromised legitimate node modifies the delivery metrics of the node to unveiling attacks in the networks.

*Li et al.* [6] present the prediction which is based on metrics abstracted from nodes contact history. In their system, nodes adopt a unique way of interpreting the contact history by making observations based on the collected encounter tickets. *Zhang et al.* [14] implemented an IDS based on Support Vector Machine (SVM) and the genetic algorithm that optimizes the parameters of SVM. The SVM examines the local traces with collected user data, system activities within a range. The IDS agents are independently responsible for attack detection if the attack is spotted these agent collaborated with neighbor nodes investigates their broader range.

*Guo et al.* [4] make use of nodes encounter records to detect or mitigate the impact of this attack. This proposed system named as misbehavior detection scheme to defend against blackhole and greyhole attacks. The previous encounters are collected and exchanged securely which can assess the trustworthiness of other nodes to detect blackhole and greyhole attacks. Evaluation is done by the method through extensive simulations using different DTN routing protocols.

*Hayajneh et al.* [11] proposed a theoretical framework that deals with collision and malicious packet dropping attacks. To avoid the packet loss, DSR-based network is applied in this framework with limited topology. However, the major concern in this work is mobility. *Yun et al.* [13] introduced a collision detection method for IEEE 802.11. This method has two phases namely, Failure Notification (FN) and Collision Notification (CN). FN analyses the false transmission using transmission time and energy time collected from the transmission history whereas CN disseminates the collision and detects the hidden stations.

*Mehdi et al.* [7] proposed a Blackhole detection method in AODV protocol. The detection algorithm checks and waits for the reply from the destination node to find the safest path for data transmission. However, it leads to the transmission delay. To avoid the transmission delay, negotiation is done with nearby nodes to have a proper route to the destination. The system achieves better accuracy with minimal additional delay and overhead.

Hence in this literature survey, it is observed that most of the resources are appropriately linked/connected, due to this problem of Black-Hole & Gray-Hole attacks in cloud computing. For this system proposed and implemented an efficient detection method against Black-Hole & Gray-Hole attacks in the cloud computing environment.

III. PROPOSED SYSTEM

The proposed detection method against Black-Hole and Gray-Hole attack in Cloud Computing is depicted in figure 3.

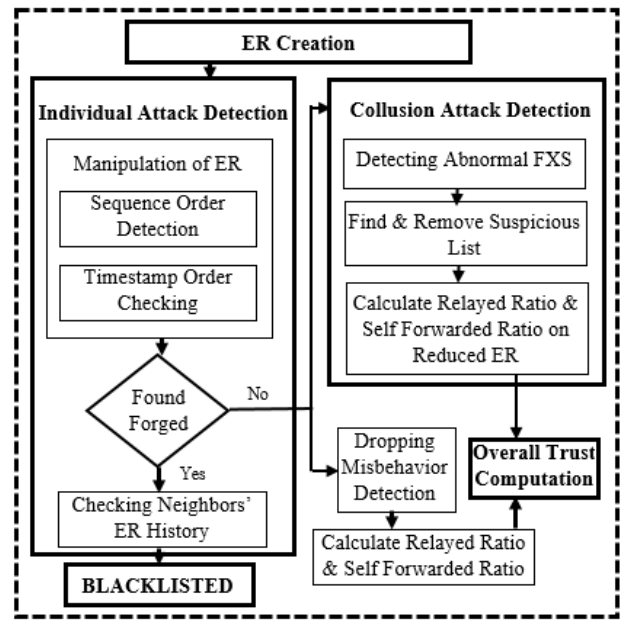

Figure 3: System Architecture

The system is integrated with individual attack and collusion attacks detection. The major phases in this proposed system are (i) Creating Encounter Records (ER) (ii) Detecting Individual Attackers (iii) Dropping Misbehavior Detection and (iv) Detecting collusion attacks. The detection process starts with ER creation whereas the communication between the sender node and receiver node starts, each node on both sides generates an ER and stores in node memory. Each ER's will have the details about the sender, receiver, timestamp, sequence number and the message details. The Detection of individual attackers can be manipulated with the help of ER. For identifying the malicious node, the major input is going to be an Encounter Records. Also, the sequence order of the ER that is generated is checked. The sequence order and timestamp order should be continuous increasing order; also some tempered has been made. The neighbors history checking can be calculated with the ER. Checking of neighbors' history can be calculated with the help of ER's. From all the above three processes, even if one violates, then it is assumed that they have tried to forge the ER; hence the packets are blacklisted.

After manipulation of ER, if the message packets are not found as forged the dropping misbehaviors detection phase is performing automatically. The attacker generates more significant portions of the messages to drop other messages. The observation of dropping misbehavior is evaluated based on the Relayed Ratio (RR) and Self-forwarded Ratio (SR). Relayed Ratio is calculated by the total number of message sent and received divided by a total number of message received and not the sent message. Consider n is a destination node; the RR of node n is defined in equation 1.

$$RR = \frac{RFM_n}{RMR_n} \quad \text{------ (1)}$$

Where the $RFM_n$ is the total number of Relayed Forwarded Message to another node by node n. $RMR_n$ represents the total number of Received Message as Relay by node n, but not forwarded to another node.

Self-Forwarding Ratio is calculated by the total number of the message generated and sent by the same user divided by the total number of message sent.

$$SR = \frac{GSM_n}{SM} \quad \text{------ (2)}$$

Where $GSM_n$ is the total number of generated and sent out messages by node n. SM denotes the total number of sent out messages. The malicious node has a lower RR and higher SR. The RR is comparably high, and SR is low in the secured node (Well behaved node). The low RR and high SR lead to the decrease the reputations.

In the cloud environment, the Black-Hole & Gray-Hole attackers are colluded with each other to camouflage their misbehavior. The Detection of Collusion Attackers is implemented to check whether they are colluding with any other malicious user to cheat the system. In this phase, the system will check the ER for a particular node where the messages are sent frequently. If the message sent ratio is high, then the node should be detected and checked for any collusion with other nodes.

The colluding attackers manipulate the ER for keeping their reputation. The strategy is making a fake ER with manipulating the RR and SR. Then the attacker changes the malicious node ER to compensate the dropped portion that is satisfied with the well-behaved threshold range. If it has multiple colluders, the attacker selects the encounter peer node for the fake record, and these peer node only create the least number of encounters in ER history. Finally, the malicious node adds the signature for the chosen colluders. To detect these situations, this system investigates the potential metrics (RR & SR) with authentic records because the colluding attackers cannot conceal the anomaly of both metrics at the same time.

Consider the node $a$ and $n$ are colluders, the node $n$ recorded as encountering $a$n in ER. The FXS metric is defined as equation 3.

$$FXS_a^n = \frac{M_a^n}{F_a^n} \quad \text{------ (3)}$$

Where $M_a^n$ is the total number of messages sent from $a$ to $n$, $F_a^n$ is the total number of encounters between $a$ and $n$ (frequency). The FXS metric identifies the fake encounters, and it should be a high abnormality ratio that differentiated from authentic ER using thresholds. Based on the FXS metric analysis colluder suspicious list are filtered from the ER, and the fake records are not met the computation process. Finally, the filter window calculates the RR & SR then verify with the threshold value. They violate the threshold, the node will be blacklisted, and overall trust computation is calculated.

IV. IMPLEMENTATION RESULTS

The CloudSim simulator is used for implementing the system design. The testbed is created with 5 servers and 50 virtual machines in mesh topological structure which can be connected and communicated with each other. The overall simulation time is 10hrs, and the message packet is generated within the range of 20-30 seconds.

To analysis of the performance of the proposed system following evaluation metric are used;
1. Precision – the truly detected malicious node infraction.

   P = TP / (TP + FP)    ------ (4)

2. Recall – It indicates the overall detection rate.

   R = TP/ (TP + FN)    ------ (5)

3. F-Score – Total number of malicious nodes are detected by a normal node in percentage.

   F-Score = 2 * (P * R) / (P + R)   ------ (6)

Where TP (True Positive) which identifies the malicious node as malicious, FP (False Positive) that indicate Non - Malicious node as Malicious node and FN (False Negative) which indicate the Malicious node as Non - Malicious node. These metrics are evaluated based on RR, SFR and threshold.

A. *Individual Attackers for Relayed Ratio*

TABLE 1

| Threshold | Precision | Recall | F-score |
|---|---|---|---|
| 0.4375 | 0.76 | 0.58 | 0.657 |
| 0.5375 | 0.781 | 0.625 | 0.694 |
| 0.5875 | 0.79 | 0.62 | 0.694 |

The table I represents the calculation of precision, Recall and F-score based on threshold values for relayed ratio on individual attackers.

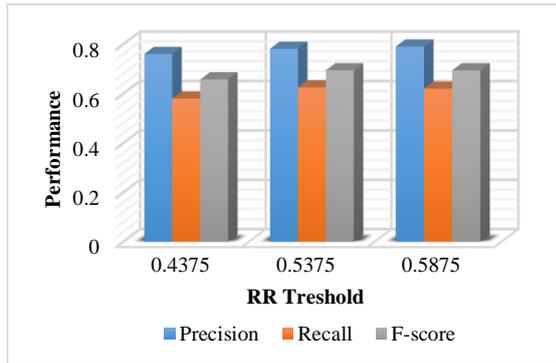

Figure 4: Relayed Ratio based on Individual Attackers

Figure 4 presents high precision with an acceptable recall ratio. Based on these values, the F-score is calculated which achieve 69% accuracy within the threshold range 0.5375 – 0.5875.

B. *Individual Attackers for Self Forwarding Ratio*

Table II represents the threshold values of self-forwarding ratio and represents the calculated evaluation metric values based on the equations 4,5,6.

TABLE II

| Threshold | Precision | Recall | F-score |
|---|---|---|---|
| 0.56 | 0.76 | 0.66 | 0.687 |
| 0.63 | 0.78 | 0.625 | 0.693 |
| 0.69 | 0.77 | 0.58 | 0.661 |

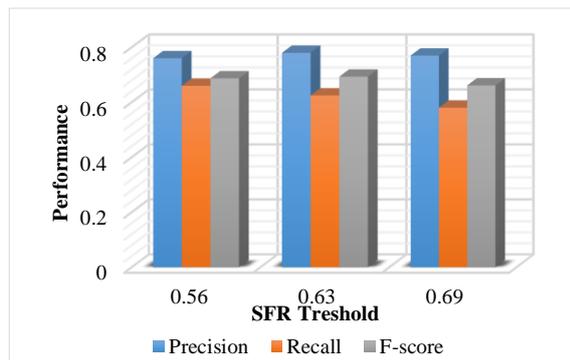

Figure 5: Self-Forwarding Ratio based on Individual Attackers

Figure 5 represents the performance which is evaluated based on the given threshold of self-forwarding ratio. The threshold range 0.56 to 0.69 gets higher precision with better recall and the 0.63 threshold achieve higher accuracy of 69%.

C. *Collusion Attackers for Relayed Ratio*

The table III represents the RR threshold range between 0.47 to 0.66 and determine the accuracy of the system based on this RR threshold for collusion attackers.

TABLE III

| Threshold | Precision | Recall | F-score |
|---|---|---|---|
| 0.47 | 0.61 | 0.9 | 0.727 |
| 0.52 | 0.63 | 0.95 | 0.757 |
| 0.66 | 0.70 | 0.953 | 0.807 |

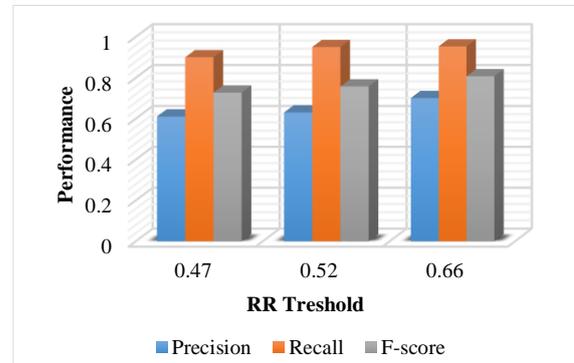

Figure 6: Relayed Ratio based on Collusion Attackers

The graphical representation of the evaluated RR threshold of the precision, recall and F-score is shown in figure 6. The graph has shown a higher detection rate and better precision in the threshold range of 0.47 to 0.66. 80% of detection accuracy is obtained in the threshold value of 0.66.

D. *Collusion Attackers for Self Forwarding Ratio*

The calculated values of precision, recall and f-score based on the SFR threshold range between 0.59 t0 0.71 represented in Table IV.

TABLE IV

| Threshold | Precision | Recall | F-score |
|---|---|---|---|
| 0.59 | 0.62 | 0.83 | 0.709 |
| 0.65 | 0.61 | 0.79 | 0.688 |
| 0.71 | 0.60 | 0.76 | 0.670 |

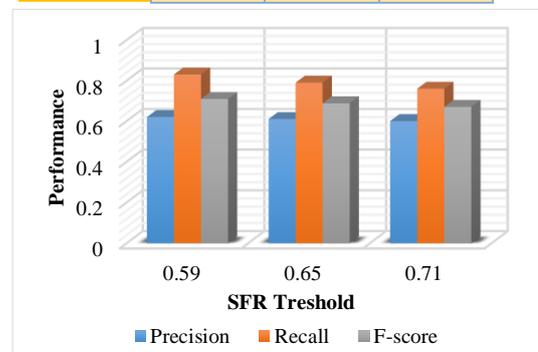

Figure 7: Self-Forwarding Ratio based on Collusion Attackers

Figure 7 presents the higher detection rate with better precision in the SFR threshold value of 0.59. This threshold value also obtains a higher accuracy of 70%.

### E. Performance of Both Individual and Collusion Attackers

Table V represents the best threshold values for both individual and collusion attackers with both relayed ratio and self -forwarding ratio. The graphical representation of both attacks detection accuracy is shown the figure 8.

TABLE V

| Threshold | Precision | Recall | F-score |
|---|---|---|---|
| Individual Attackers | 0.781 | 0.625 | 0.69 |
| Collusion Attackers | 0.70 | 0.95 | 0.80 |

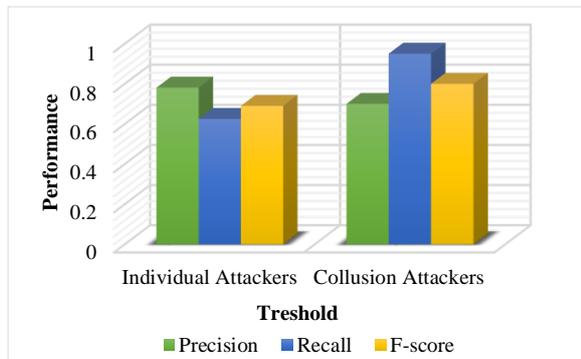

Figure 8: The ratio for Individual and Collusion Attackers

## V. CONCLUSION AND FUTURE WORK

Blackhole and greyhole assaults are a thoughtful menace to the cloud computing environment. In the prevailing system they protect beside packet dropping assaults in the cloud but furthermost of them nosedive to thwart the collusion of malicious nodes. The proposed detection method can magnificently thwart specific attackers, and the system achieves colluding malevolent nodes with high detection rate and low false positive rate when fluctuating the number of colluding nodes and with a wide range of packet dropping probability. Colluding attackers are not found early due to forged ER. So to prevent colluding attackers this system increases and decreases the reputation value of the threshold obtained. The performance is evaluated based on the relational ratio, and self-forwarding ratio with their predicated threshold ranges and the system achieve higher accuracy of 80%.

REFERENCES

[1] Chuah. M, Yang. P, and Han. J, "A ferry-based intrusion detection scheme for sparsely connected ad hoc networks" in *Proc. 4th Annu. Int. Conf. Workshop Security Emerging Ubiquitous Computer*, pp. 1–8, 2007.

[2] Divyasree I R and Selvamani K, "Defeating the Distributed Denial of Service Attack in Cloud environment. A survey", in *International Conference on Circuit, Power and Computing Technologies (ICCPCT), IEEE*, pp.1-8, 2017.

[3] Fan-Hsun Tseng, Li-Der Chou and Han-Chieh Chao, "A survey of black hole attacks in wireless mobile ad hoc networks", in *Human-centric Computing and Information Sciences*, Vol.1, No.4, 2011.

[4] Guo. Y, Schildt. S, and Wolf. L, "Detecting blackhole and greyhole attacks in vehicular delay tolerant networks", in *Proc. IEEE 5th Int. Conf. Commun. Syst. Netw.*, pp. 1–7, 2013.

[5] Khalil I. M, Khreishan A and Azeem, M, "Cloud computing security: A survey", *Computers,* vol.3, pp-1-35, 2014.

[6] Li. F, Wu. J, and Srinivasan. A, "Thwarting blackhole attacks in disrupt-tolerant networks using encounter tickets", in *Proc. INFOCOMM*, pp. 2428–2436, 2009.

[7] Mehdi Medadian and Khossro Fardad, "Proposing a Method to Detect Black Hole Attacks in AODV Routing Protocol", in *European Journal of Scientific Research*, ISSN 1450-216X Vol.69, No.1, pp.91-101, 2012.

[8] Ren. Y, Chuah. M, J. Yang, and Chen. Y, "MUTON: Detecting malicious nodes in disrupt-tolerant networks" in *Proc. IEEE Wireless Commun. Netw. Conf.*, pp. 1–6, 2010.

[9] Riasudheen.H and Selvamani.k, "A Survey on Backup Routing Schemes in MANETs", in *International Journal of emerging technology in computer science & Electronics*, Vol.24, Issue.6, pp. 27-31,2017.

[10] Sivagurunathan. S and Prathapchandran. K, "Trust Based Security Model to Withstand Against Black Hole and Grey Hole Attacks in Military Based Mobile Ad Hoc Networks", in *International Journal of Mobile Network Communications & Telematics (IJMNCT),* Vol. 6, No.1, February 2016.

[11] Thaier Hayajneh, Prashant Krishnamurthy, David Tipper, and Taehoon Kim, "Detecting Malicious Packet Dropping in the Presence of Collisions and Channel Errors in Wireless Ad hoc Networks ", in *IEEE International Conference on Communications,* pp.1-6, 2009.

[12] Thi Ngoc Diep Pham and Chai Kiat Yeo, "Detecting Colluding Blackhole and Greyhole Attacks in Delay Tolerant Networks", in *IEEE Transactions on Mobile Computing*, Vol.15, No. 5, pp.1116-1129, 2016.

[13] Yun J.H. and S.W. Seo, "Novel collision detection scheme and its applications for IEEE 802.11 wireless lans," *Computer Communications*, Vol. 30, No. 6, pp. 1350–1366, 2007.

[14] Zhang Chunfei and Fanf Zhiyi, "A new Distributed Intrusion Detection system model based on SVM in Wireless Mesh Networks", in *Journal of Information & computational Science*, Vol. 12, No. 2, pp. 751-759, 2015.